\documentclass[aps,prb,reprint,showpacs]{revtex4-1}
\usepackage[dvipdfmx]{graphicx}

\usepackage{amsmath,amssymb}
\usepackage{color}

\usepackage{bm}
\def\vb#1{{\bm#1}}
\def\v#1{\mathbf{#1}}			

\def\vB{\v{B}}

\def\vv{\v{v}}

\def\r{\v{r}} 					
\def\p{\v{p}} 					
\def\k{\v{k}} 					

\def\vA{\v{A}}

\def\vB{\v{B}}

\def\vOmega{\vb{\Omega}}
\def\vomega{\vb{\omega}}

\def\vpi{\vb{\pi}}
\def\vsigma{\vb{\sigma}}

\def\vSigma{\vb{\Sigma}}

\def\vv{\v{v}}

\def\del{\partial}

\begin{document}


\title{Theory of spin hydrodynamic generation}



\author{M. Matsuo$^{1,2}$, Y. Ohnuma$^2$, and S. Maekawa$^2$}
\affiliation{%
$^1$Advanced Institute for Materials Research, Tohoku University, Sendai 980-8577, Japan\\
$^2$Advanced Science Research Center, Japan Atomic Energy Agency, Tokai, Ibaraki 319-1195, Japan.
}%


\date{\today}

\begin{abstract}
Spin-current generation by fluid motion is theoretically investigated. 
Based on quantum kinetic theory, the spin-diffusion equation coupled with fluid vorticity is derived. 
We show that spin currents are generated by the vorticity gradient in both laminar and turbulent flows and that  
the generated spin currents can be detected by the inverse spin Hall voltage measurements, which are predicted to be proportional to the flow velocity in a laminar flow. 
In contrast, the voltage in a turbulent flow is proportional to the square of the flow velocity. 
This study will pave the way to fluid spintronics. 

\end{abstract}


\maketitle 

\section{Introduction}
Spintronics is 
an emerging field in condensed matter physics, which focuses on the generation, manipulation, and detection of spin current\cite{MaekawaEd2012}. 
Two mechanisms of spin-current generation have been repeatedly confirmed: the  spin-orbit coupling driven and the exchange coupling driven mechanisms.
The spin-Hall effect\cite{Valenzuela2006,Saitoh2006,NLSV} belongs to the former class as it relies on the spin-orbit scattering. 
A typical example of the latter class is the spin pumping\cite{Tserkovnyak2002,Uchida2008,UchidaASP}, which originates from the dynamical torque-transfer process during magnetization due to nonequilibrium spin accumulation. 

Recently, an alternative scheme has been proposed, wherein spin-rotation coupling\cite{SRC} is exploited for generating spin currents\cite{MSOI-SC,SAW-SC:Matsuo,SAW-SC:Hamada}. The spin-rotation coupling refers to the fundamental coupling between spin and mechanical rotational motion and emerges in both ferromagnetic\cite{Barnett1915} and paramagnetic\cite{Ono2015,Ogata2017} metals as well as in nuclear spin systems\cite{Chudo2014,Chudo2015}. This coupling allows the interconversion of spin and mechanical angular momentum. 

Spin-current generation has been experimentally demonstrated using the mechanical rotation of a liquid metal\cite{Takahashi2016}. In the experiment, the induced mechanical rotation in a turbulent pipe flow of Hg and Ga alloys  is utilized to generate the spin current.  

In this paper, we theoretically investigate the fluid-mechanical generation of spin current in both laminar and turbulent flows of a liquid metal and predict that 
the fluid velocity dependence of the spin current under laminar conditions will be qualitatively different from that in the turbulent flow. 
First, we show that the spin-vorticity coupling emerges in a liquid metal and derive the spin-diffusion equation in a liquid-metal flow based on quantum kinetic theory. 
We solve the spin-diffusion equation and reveal that the spin current is generated by the vorticity gradient. By solving the equation under both laminar- and turbulent-flow conditions, 
 the inverse spin Hall voltage in the laminar liquid flow is predicted to be linearly proportional to the flow velocity, whereas the voltage in the turbulent flow is proportional to the square of the flow velocity. 
 Our study will pave the way to fluid spintronics, where spin and fluid motion are harmonized.

\section{Spin vorticity coupling\label{Sec:SpinVorticity}}
To consider the inertial effect on an electron due to nonuniform acceleration,   
we begin with the generally covariant Dirac equation\cite{Bib:SpinConnection}, which governs the fundamental theory for a spin-1/2 particle in a curved space-time:
\begin{eqnarray}
\left[i \gamma^{\mu}  \left(p_{\mu }-qA_{\mu} -i\hbar \Gamma_{\mu} \right)  +mc  \right] \Psi =0,\label{gDirac}
\end{eqnarray}
where $c, \hbar, q=-e,$ and $ m$ represent the speed of light, the Planck's constant, the charge of an electron, and the mass of an electron, respectively.  
Equation (\ref{gDirac}) includes
two types  of gauge potentials:  
the U(1) gauge potential, $A_{\mu}$, and the spin connection, $\Gamma_{\mu}$.
The former originates from external electromagnetic fields 
and the latter describes gravitational and inertial effects upon electron charge and spin.
The spin connection, $\Gamma_{\mu}$, is determined by the metric $g_{\mu \nu}(x)$. 
The coordinate-dependent Clifford algebra can be expressed by $\gamma^{\mu}=\gamma^{\mu}(x)$, and it satisfies $\{ \gamma^{\mu}(x), \gamma^{\nu}(x) \}=2g^{\mu \nu}(x)\, (\mu,\nu=0,1,2,3)$ with the inverse metric given by $g^{\mu\nu}(x)$.

In the following, we focus on a single electron in 
a conductive viscous fluid.
The motion of the viscous fluid is effectively described by its flow velocity, $\vv (x)$, 
which is the source of the gauge potential on an electron, $\Gamma_{\mu}$, 
and reproduces inertial effects on the electron charge and spin, as explained below. 
We assume that the flow velocity is much less than the speed of light, $|\vv| \ll c$. 
The coordinate transformation from a local rest frame of the fluid to an inertial frame is written as
$d\v{r}'=d\v{r} + \vv(x) dt,$
and the space-time line element in the local rest frame is
\begin{eqnarray}
ds^{2}&=&g_{\mu\nu}dx^{\mu}dx^{\nu} \nonumber \\
&=&[-c^{2} + \vv^{2} ] dt^{2} + 2 \vv \cdot d\r dt +d\r^{2}.
\end{eqnarray}
 Then, the metric becomes 
 \begin{eqnarray}
g_{00}=-1+v^{2}/c^{2}, g_{0i}=g_{i0}=v_{i}/c, g_{ij}=\delta_{ij}. \label{metric}
\end{eqnarray}
Equations (\ref{gDirac}) and (\ref{metric}) lead to the Dirac Hamiltonian in the local rest frame:
\begin{eqnarray}
H &=& \beta mc^{2} + c \vb{\alpha} \cdot  \vb{\pi} +qA_{0} -\frac12 q \v{A}\cdot \vv - \frac12 \{ \vv, \vb{\pi} \}
	-\frac12 \vb{\Sigma} \cdot \vb{\omega}. \nonumber \\ \label{Hlr}
\end{eqnarray}
Here $ \beta$ and $ \vb{\alpha}$ are the Dirac matrices and $\vb{\Sigma} $ is the spin operator. 
Moreover, $\vb{\pi}=\p - q\v{A} $ refers to the mechanical momentum, $\vomega = \nabla \times \vv$ is the vorticity of the fluid, and $\{ \vv, \vb{\pi} \}=\vv \vb{\pi}+ \vb{\pi}\vv$. 
Equation (\ref{Hlr}) is a generalization of the Dirac equation in a rigidly rotating frame. If the velocity is chosen to be $\vv(x) = \vOmega \times \r$ with a constant rotation frequency, $\vOmega$, 
then the fourth term $\{ \vv, \vb{\pi}\}/2 $ is a representative of the coupling of the rotation and the orbital angular momentum, $-\vOmega \cdot (\r \times \vpi)$, which reproduces quantum-mechanical versions of the Coriolis, centrifugal, and Euler forces, as shown below.  
The fifth term, $-\vb{\Sigma} \cdot \vb{\omega}/2$, can be called the ``spin-vorticity coupling,'' which reproduces the spin-rotation coupling $-\vSigma \cdot \vOmega$ because the vorticity $\vomega$ is reduced to the rotation frequency $\vOmega$ as $\vomega = 2 \vOmega$ for rigid motion. 
Thus, Eq. (\ref{Hlr}) reproduces the Dirac equation in the rotating frame.

\section{Inertial forces on an electron due to viscous-fluid motion}
Using the lowest order of the Foldy-Wouthuysen-Tani expansion\cite{FWT} for Eq. (\ref{Hlr}), we obtain the Schr\"odinger equation for an electron's two-spinor wave function, $\psi$, in the fluid:
\begin{eqnarray}
&&i\hbar \frac{\del \psi}{\del t}=H \psi,  \nonumber  \\
H &=& \frac{1}{2m}\vb{\pi}^{2} +qA_{0} - \mu_{B} \vsigma \cdot \vB \nonumber \\
&&-\frac12 q\v{A}\cdot \vv - \frac12 \{\vv, \vb{\pi} \}
	-\frac12 \v{S} \cdot \vb{\omega}, \label{Hsv}
\end{eqnarray}
with $\mu_{B}=q\hbar/2m$, $\vB=\nabla \times \vA$, and $\v{S}=(\hbar/2)\vsigma$.

From Eq. (\ref{Hsv}), the Heisenberg equation for an electron in the fluid is obtained as
\begin{eqnarray}
&&\dot{\v{r}} = \frac{1}{i\hbar}[\v{r}, H]=\frac{\vb{\pi}}{m} + \v{v} \\
&&m \ddot{\v{r}} = \frac{1}{i\hbar}[m\dot{\v{r}},H] + m\frac{\del \v{v}}{\del t} = \v{F}
\end{eqnarray}
where the operator $F_{i}$ whose expectation value corresponds to a semi-classical force is given by
\begin{eqnarray}
&&\v{F}= \v{F}_{\rm em} + \v{F}_{\rm c} + \v{F}_{\sigma},\\
&&\v{F}_{\rm em} = q \left( \v{E}+ (\dot{\r} -\vv)\times \vB  \right), \label{Fem}\\
&&\v{F}_{\rm c} = m(\dot{\r}\cdot \nabla)\vv - (\nabla v_{i})\dot{r}_{i} + (\nabla v_{i})v_{i} + m\frac{\del \vv}{\del t},\label{Fc} \\
&&\v{F}_{\sigma} = \nabla \left\{ \mu_{B} \vsigma \cdot  \left(\vB +\frac{\vomega}{2\gamma}  \right)  \right\}.\label{Fs}
\end{eqnarray}
Equation (\ref{Fem}) represents the electromagnetic force in a conductive viscous fluid. In the case of a rigid rotation, $\vv(x)= \vOmega \times \r$, the first and second terms in Eq. (\ref{Fc}) reproduce the Coriolis force, $- 2m\dot{\r} \times \vOmega$, the third term becomes the centrifugal force, $m \vOmega \times (\vOmega \times \r) $, and the last term corresponds to the Euler force.

Equation (\ref{Fs}) is an expression for the Stern--Gerlach force, which originates from the gradient of the combination of the Zeeman term, $\mu_{B}\vsigma \cdot \vB$, and the spin-vorticity coupling term, $\hbar \vsigma \cdot \vomega/2$: 
\begin{eqnarray}
H_{\vsigma}=-\mu_{B} \vsigma \cdot \left(\vB + \frac{\vomega}{2\gamma} \right), \label{Hsigma}
\end{eqnarray}
where $\gamma = gq/2m$ is the gyromagnetic ratio with $g=2$. 
This indicates that the inertial effect due to fluid motion is equivalent to the effective magnetic field $B_{\omega}=\gamma^{-1} \vomega/2$.
In the following paragraphs, we demonstrate that the effective field is crucial for generating  the spin current.

\section{Spin-diffusion equation in a liquid-metal flow}
\begin{figure}[!hbtp]
\begin{center}
\includegraphics[scale=0.4]{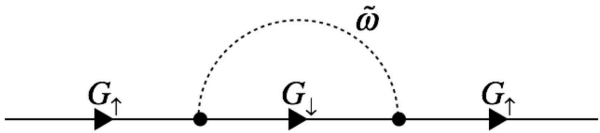}
\caption{Contribution to the self-energy $\Sigma$ originating from the spin-vorticity coupling. }\label{fey-graph}
\end{center}
\end{figure}

To investigate spin-current generation due to the spin-vorticity coupling, we derive the spin-diffusion equation by using quantum kinetic theory.  
Starting with the quantum kinetic equation:
\begin{eqnarray}
	\frac{\del G^<}{\del t} - \frac{1}{\hbar}&& \frac{\del \mbox{Re}\Sigma^R}{\del R} \frac{\del G^<}{\del k} + v_k \frac{\del G^<}{\del R} \nonumber\\
	&&=\frac{1}{\hbar}(G^K \mbox{Im} \Sigma^R - \mbox{Im} G^R \Sigma^K),
\end{eqnarray}
where $G$ is the nonequilibrium Green's function of an electron, $\Sigma$ is the self-energy of the electron, and $v_k$ is the group velocity of the electron. We consider the effects of the impurity potential, the spin-orbit potential, and the spin-vorticity coupling:
\begin{eqnarray}
	H_{\rm int} = V_{\rm imp} + \eta_{\rm so} \vsigma \cdot (\nabla V_{\rm imp} \times \v{p}) - \frac{\hbar}4 \vsigma \cdot \vomega, 
\end{eqnarray}
where $V_{\rm imp}$ is an ordinary impurity potential and $\eta_{\rm so}$ is the spin-orbit coupling parameter.
Using a quasi-particle approximation, the quantum kinetic equation reduces to the spin-dependent kinetic equation:
\begin{eqnarray}
	\frac{\del f^\sigma_{ktr}}{\del t}-\frac{1}{\hbar}\frac{\del \Sigma^{\sigma,R}_{k\varepsilon}}{\del R} \frac{\del f^{\sigma}_{krt}}{\del k} + v_k \frac{\del f^\sigma_{krt}}{\del R} \nonumber\\
	=-\frac{f^\sigma_{ktr}-f^0_k}{\tau_{\rm imp}} -\frac{f^\sigma_{ktr}-f^{-\sigma}_{ktr}}{\tilde\tau_{\rm sf}}, \label{Boltzmann}
\end{eqnarray}
where $f^\sigma_{krt}$  is the distribution function of an electron with spin $\sigma$, $f^0$ is the equilibrium distribution function of an electron, and 
$\tau_{\rm imp}$ is the transport-relaxation time given by
\begin{eqnarray}
\tau^{-1}_{\rm imp} = 2\pi n_{\rm imp} D_{\rm F}V^2_{\rm imp}/\hbar	
\end{eqnarray}
with the impurity density $n_{\rm imp}$.
The spin-flip relaxation time $\tilde\tau_{\rm sf}$ given by
\begin{eqnarray}
\tilde\tau^{-1}_{\rm sf}(k) = \tau^{-1}_{\rm sf}(k) +  \tau^{-1}_{\rm sv}(k),	
\end{eqnarray} 
where the spin-life time due to the spin-orbit coupling is 
\begin{eqnarray}
\tau^{-1}_{\rm sf}=2\eta^{-2}_{\rm so}\tau^{-1}_{\rm imp}	
\end{eqnarray}
and the spin-life time due to the spin-vorticity coupling $\tau_{\rm sv}$ is given by 
\begin{eqnarray}
\tau^{-1}_{\rm sv}(\r,\k,t,\omega)=D_{\rm F}\tilde{\omega}(\r,\k,t,\omega), 	
\end{eqnarray}
where $\tilde{\omega}(\r,\k,t,\omega)$ is the Wigner representation of the kinetic component of the two particle correlation function defined by
\begin{eqnarray}
\tilde{\omega}(\r,\k,t,\omega)\equiv \int_{\delta \r \delta t}\!\!\!\! \tilde{\omega}(\r,\delta \r,t,\delta t) e^{i(\k\cdot \delta \r - \omega \delta t)}	
\end{eqnarray}
with 
\begin{eqnarray}
&&\tilde{\omega}(\r,\delta \r,t,\delta t) \nonumber \\
&&= {\rm Tr}\Big[\hat\rho \omega^+\Big(\r- \frac{\delta\r}{2},t-\frac{\delta t}{2} \Big) \omega^-\Big(\r+\frac{\delta \r}{2},t+\frac{\delta t}{2} \Big) \Big].	
\end{eqnarray}
Here $\hat\rho$ is the density matrix of the fluid (Fig. \ref{fey-graph}).
Using the expansion: 
\begin{eqnarray}
f^\sigma_{krt} = f^0_k + \del_{E_k}f^0_k (\sigma\delta \mu + \hbar \omega_{krt}/2),	
\end{eqnarray}
the momentum average of the kinetic equation is reduced to the generalized spin-diffusion equation:
\begin{eqnarray}
(\del_t -D_s \del_x^2 + \tilde\tau_{\rm sf}(k_F)^{-1}) \delta \mu_S =  -\frac{\hbar}{\tilde\tau_{\rm sf}(k_F)} \zeta \omega^z,\label{g.spin-diff}
\end{eqnarray}
where $D_s$ is the diffusion constant and $\zeta$ is the renormalization factor of the spin-vorticity coupling defined by
\begin{eqnarray}
 \zeta =\frac{\int_0^{k_F} dk [\del_k f^0_k \omega^z_{rkt}\tilde\tau_{\rm sf}(k)^{-1}]}{ \omega^z(r,t)\tilde{\tau}_{\rm sf}(k_F)^{-1}\int_0^{k_F} dk [\del_k f^0_k] } 	
\end{eqnarray}
with the Fermi wave number $k_F$. 
Based on the non-equilibrium Green's function method, the renormalization factor is found to depend on the microscopic parameters including the transport-relaxation time, the spin-flip life time, resulting from impurity scatterings and an extrinsic spin-orbit coupling.

In nonequilibrium steady-state conditions, this equation may be further reduced to
\begin{eqnarray}
\Big(\nabla^2-\frac{1}{\lambda^2} \Big) \delta \mu_s=  \frac{\hbar \zeta \omega}{\lambda^2},\label{sSD}
\end{eqnarray}
where $\lambda$ denotes the spin-diffusion length.

\section{Spin current from fluid motion}
\subsection{Spin current from laminar flow between plates}
We now solve the spin-diffusion equation under a typical laminar flow condition.  
The equations of motion for an incompressible viscous fluid are well described by the Navier-Stokes (NS) equation:
\begin{eqnarray}
\frac{\del \vv}{\del t} + (\vv \cdot \nabla) \vv = - \frac{1}{\rho} \nabla p + \frac{\eta}{\rho} \nabla^{2} \vv, \label{NV}
\end{eqnarray}
where $\rho$ is the fluid density, $\eta$ is the viscosity coefficient, and $p$ is the pressure.   
In the following derivation, we use a solution of the NS equation. Moreover, the vorticity field calculated from the solution is inserted into the static spin-diffusion equation in  (\ref{sSD}) to obtain the generated spin current.

\begin{figure}[!hbtp]
\begin{center}
\includegraphics[scale=0.4]{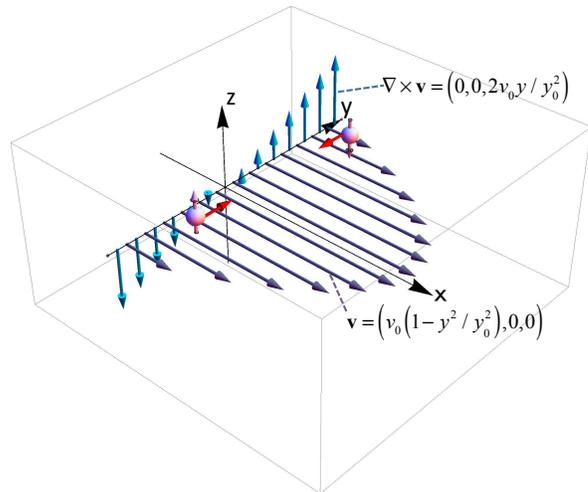}
\caption{Representation of the spin current in the two-dimensional Poiseuille flow. 
The parallel flow between the two parallel planes, $y=\pm y_{0}$, creates the velocity field $\v{v}=(v_{0}(1-y^{2}/y_{0}^{2}),0,0)$. The vorticity of the flow emerges in the $z$-direction: $\nabla \times \v{v} = (0,0,2v_{0}y/y_{0}^{2})$. Then, the gradient of the vorticity generates the $z$-polarized spin-current in the $y$-dicrection.  
}\label{figP}
\end{center}
\end{figure}

We consider the parallel flow enclosed between two parallel planes with a distance of $2y_{0}$ as shown in Fig. \ref{figP}.
The solution to Eq. (\ref{NV}) is the well-known two-dimensional Poiseuille flow\cite{LandauFluid}:
\begin{eqnarray}
v_{x}=v_{0} \{1-\left(y/y_{0} \right)^{2} \}, v_{y}=v_{z}=0,  \label{2dimP}
\end{eqnarray}
where
\begin{eqnarray}
\frac{v_{0}}{y_{0}^{2}}=-\frac{1}{2\eta} \frac{dp}{dx}.
\end{eqnarray}
In this case, 
the vorticity becomes 
\begin{eqnarray}
	\vomega = \nabla \times \vv = (0,0,2v_0 y/y_0^2 ).\label{vomega-2dimP}
\end{eqnarray}
Inserting Eq. (\ref{vomega-2dimP}) into Eq. (\ref{sSD}), we obtain the $z$-polarized spin current as
\begin{eqnarray}
J_{s,y}^z &&= \frac{\sigma_0}{e} \frac{\del}{\del y} \delta \mu_s^z (y) 
= 2\zeta \frac{\hbar \sigma_0}{e} \frac{v_0}{y_0^2}  \Big[ 1 -\frac{\cosh (y/\lambda)}{ \cosh (y_0/\lambda)} \Big] \nonumber\\
&& \approx 2\zeta \frac{\hbar \sigma_0}{e} \frac{v_0}{y_0^2},\label{Js-Po}
\end{eqnarray} 
when $y_0 \gg \lambda$.

\subsection{Spin current from laminar flow in a pipe.}
\begin{figure}[htbp]
\begin{center}
\includegraphics[scale=0.4]{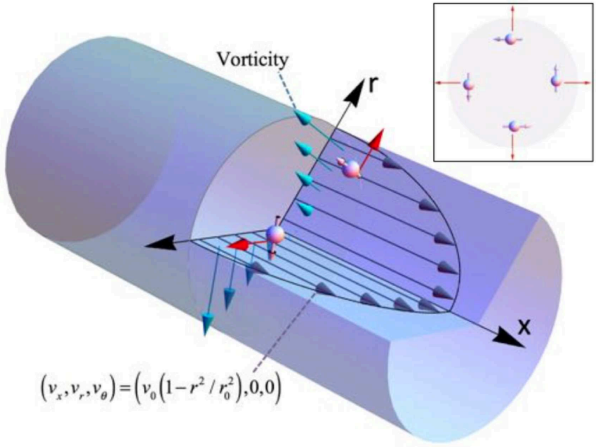}
\end{center}
\caption{Representation of the spin current for the Hagen--Poiseuille flow. A steady viscous flow in a pipe of circular cross-section with radius $r_{0}$ creates the velocity field, $(v_{x},v_{r},v_{\theta})=(v_{0}(1-r^{2}/r_{0}^{2}),0,0)$, in the cylindrical coordinate $(x,r,\theta)$. In this case, the vorticity gradient generates the $\theta$-polarized spin current in the radial direction. The inset shows the cross section of the pipe.   }
\label{figHP}
\end{figure}

Let us consider a steady flow in a pipe of circular cross-section with radius $r_{0}$ (Fig. \ref{figHP}). 
In this case, the solution to Eq. (\ref{NV}) is the Hagen--Poiseuille flow\cite{LandauFluid}:
\begin{eqnarray}
v_{x}=v_{0} \{1-\left(r/r_{0} \right)^{2} \}, v_{r}=v_{\theta}=0,  
\end{eqnarray}
when
\begin{eqnarray}
\frac{v_{0}}{r_{0}^{2}}=-\frac{1}{4 \eta} \frac{dp}{dx}.
\end{eqnarray}
The $\theta$-polarized spin current, which flows in the radial direction, is given by
\begin{eqnarray}
J_{s,r}^{\theta} \approx  2\zeta \frac{\hbar \sigma_0}{e} \frac{v_0}{r_0^2}.\label{Js-HP}
\end{eqnarray}

\subsection{Spin current from turbulent flow in a pipe}
We also consider a turbulent flow in the pipe. 
Velocity distribution in a turbulent flow in a pipe is well described as\cite{LandauFluid}
\begin{eqnarray}
\frac{v_x (r)}{v_*} = 
 \left\{ \begin{array}{ll}
  	\frac{v_* (r_{0}-r)}{\nu} 				& (r_0 -\delta_0 < r < r_0 ) \\
   \frac{1}{\kappa} \ln \frac{v_* (r_{0}-r)}{\nu} + A 	& (
0< r < r_0 -\delta_0 )
  \end{array} \right.
\end{eqnarray}
where $v_*$ is the friction velocity, $r_{0}$ is the internal radius of the pipe, $\nu$ is the kinetic viscosity, $\kappa$ is the Karman constant, $A=5.5$ for the mercury, and $\delta_0$ is the thickness of the viscous sublayer. 
The friction velocity is related to the velocity distribution $v_x(r)$ as
$v_* = \sqrt{\nu \Big| \frac{\del v_x}{\del r}}\Big|_{r=r_0}$.
The region near the inner wall $(r_0 -\delta_0 < r < r_0 )$ is called the viscous sublayer. 

In the cylindrical coordinate $(x, r, \theta)$ (Fig. \ref{figHP}), the vorticity, $\omega_{\theta} (r) = -\del_r v_x (r)$, is given by
\begin{eqnarray}
\omega_{\theta} (r) = 
 \left\{ \begin{array}{ll}
  	\frac{v_*^2 }{\nu} 				& (r_0 -\delta_0 < r < r_0) \\
   \frac{v_*}{\kappa} \frac{1}{r_0 -r} 	& (
0< r < r_0 -\delta_0 ) \label{vorticity}
  \end{array} \right.
\end{eqnarray}
The spin current is generated mostly near the viscous sublayer, especially around $r\approx r_o - \delta_0$, where the vorticity gradient is the largest. 
Then, the spin current becomes 
\begin{eqnarray}
J_{s,r}^{\theta} \approx  2\zeta \frac{\hbar \sigma_0}{e}  \frac{v_*}{\kappa (r-r_{0})^{2}}.\label{Js-Turb}
\end{eqnarray}

\subsection{Inverse spin Hall voltage}
Finally, we investigate the inverse spin Hall voltage owing to the spin-current generation under the laminar and turbulent flow conditions. 

Following the voltage measurement by Takahashi et al.\cite{Takahashi2016}, we consider the inverse spin Hall voltage to be parallel to the flow velocity (the $x$-direction). The spin current is then converted into the electric voltage because of the spin-orbit coupling in the liquid metal and can be expressed as 
\begin{eqnarray}
	V_{\rm ISHE}^{\rm Lam} = \frac{L}{\sigma_0}\frac{2e}{\hbar}\theta_{\rm SHE} J_s
\end{eqnarray}
where $V_{\rm ISHE}$ is the inverse spin Hall voltage, $L$ is the length of the channel, $\theta_{\rm SHE}$ is the spin Hall angle of the liquid metal, and $J_s$ represents the generated spin current: $J_{s,y}^z = J_s^z$ or $J_{s,r}^\theta$. In the case of the Hagen--Poiseuille flow, the voltage is given by
\begin{eqnarray}
	V_{\rm ISHE}^{\rm Lam} = 2\zeta \theta_{\rm SHE} \frac{\hbar}{e} \frac{L }{y_0^2}v_0.
\end{eqnarray}
This indicates that the generated voltage in a laminar flow is proportional to the flow velocity $v_0$. 

Contrast to the laminar flow case, 
the voltage in a turbulent flow proportional to the square of the flow velocity:
\begin{eqnarray}
V_{\rm ISHE}^{\rm Turb} &= &
\frac{\theta_{\rm SHE} L}{\sigma_0 \pi r_0^2} 
\times  \Big(  \int_0^{r_0-\delta_0} + \int_{r_0-\delta_0}^{r_0}\Big)  2\pi r dr J_{s,r}^\theta   \nonumber \\
&\approx& \zeta \theta_{\rm SHE} \frac{4L}{r_0}\frac{\hbar}{e} \frac{1}{\kappa \nu \mathcal{R}e^\delta} v_*^{\,\,2},
\end{eqnarray}
where $\mathcal{R}e^\delta = \delta_0 v_*/\nu $ is the Reynolds number defined by the friction velocity. 

Making use of the material parameter values for the turbulent condition of the mercury\cite{Takahashi2016}, $\kappa = 1.2 \times 10^{-7}$,  $\nu=1.2 \times 10^{-7} {\rm m}^2{\rm s}^{-1}$, $L=400\times 10^{-3}$ m, $r_0 = 0.2 \times 10^{-3}$ m, $v_* = 0.1$ m/s and $V_{\rm ISHE}^{\rm Turb}=100 \times$ nV, we obtain $\zeta \theta_{\rm SHE} = 1.1$. 
Taking $\theta_{\rm SHE} = 10^{-2}$ as an example, we find the renormalization factor $\zeta$ to be $10^{2}$. 

Furthermore, we estimate the voltage in the Hagen--Poiseuille flow. 
Although the renormalization factor $\zeta$ under a laminar-flow condition is generally different from that under a turbulent condition, we assume that the factor in a laminar flow is the same order of that in the turbulent flow as $\zeta \theta_{\rm SHE} \approx 1$. 
Then choosing $L=80$mm and $r_0=0.1$mm, the computed inverse spin Hall voltage is $V_{\rm ISHE}\approx 4$ nV. 

\section{Conclusion}%
In this paper, we have investigated spin-current generation due to fluid motion. 
The spin-vorticity coupling was obtained from the low energy expansion of the Dirac equation in the fluid. Owing to the coupling, the fluid vorticity field acts on electron spins as an effective magnetic field. 
We have derived the generalized spin-diffusion equation in the presence of the effective field based on the quantum kinetic theory. 
Moreover, we have evaluated the spin current generated under both laminar- and turbulent-flow conditions, including the Poiseuille and Hagen--Poiseuille flow scenarios, and the turbulent flow in a fine pipe. The generated inverse spin Hall voltage is linearly proportional to the flow velocity, whereas that in a turbulent-flow environment is proportional to the square of the velocity. 
Our theory proposed here will bridge the gap between spintronics and fluid physics, and pave the way to fluid spintronics.

\begin{acknowledgements}
The authors are grateful to R. Takahashi, K. Harii, H. Chudo, E. Saitoh and J. Ieda for valuable comments.
This work is financially supported by ERATO, JST, 
Grant-in-Aid for Scientific Research on Innovative Areas “Nano Spin Conversion Science” (26103005),
Grant-in-Aid for Scientific Research C (15K05153), 
Grant-in-Aid for Scientific Research B (16H04023), 
and Grant-in-Aid for Scientific Research A (26247063) 
from MEXT, Japan.	
\end{acknowledgements}



\begin{thebibliography}{10}

\bibitem{MaekawaEd2012}S. Maekawa, S. Valenzuela, E. Saitoh, and T. Kimura ed., {\it Spin Current} (Oxford University Press, Oxford, 2012).

\bibitem{Valenzuela2006}S. O. Valenzuela and M. Tinkham, Nature (London) {\bf 442}, 176 (2006).

\bibitem{Saitoh2006}E. Saitoh, M. Ueda, H. Miyajima, and G. Tatara. Appl. Phys. Lett. {\bf 88}, 182509 (2006).
\bibitem{NLSV}T. Kimura, Y. Otani, T. Sato, S. Takahashi, and S. Maekawa, Phys. Rev. Lett. {\bf 98}, 156601 (2007); L. Vila, T. Kimura, and Y. Otani, Phys. Rev. Lett. {\bf 99} 226604 (2007).

\bibitem{Tserkovnyak2002}Y. Tserkovnyak, A. Brataas, and G. E. W. Bauer, Phys. Rev. Lett. {\bf 88}, 117601 (2002). 

\bibitem{Uchida2008}K. Uchida, S. Takahashi, K. Harii, J. Ieda, W. Koshibae,
K. Ando, S. Maekawa, and E. Saitoh, Nature (London) {\bf 455}, 778
(2008).
\bibitem{UchidaASP}
K. Uchida, H. Adachi, T. An, T. Ota, M. Toda, B. Hillebrands, S. Maekawa, and E. Saitoh, Nature Mater. {\bf 10}, 737 (2011).

\bibitem{SRC}C. G. de Oliveira and J. Tiomno, Nuovo Cimento {\bf 24}, 672 (1962); B. Mashhoon, Phys. Rev. Lett. {\bf 61}, 2639 (1988); F. W. Hehl and W.-T. Ni, Phys. Rev. D {\bf 42}, 2045 (1990).
\bibitem{MSOI-SC}M. Matsuo, J. Ieda, E. Saitoh, and S. Maekawa, Phys. Rev. Lett. {\bf 106}, 076601 (2011); M. Matsuo, J. Ieda, E. Saitoh, and S. Maekawa, Appl. Phys. Lett. {\bf 98}, 242501 (2011); M. Matsuo, J. Ieda, E. Saitoh, and S. Maekawa, Phys. Rev. B{\bf 84}, 104410 (2011).

\bibitem{SAW-SC:Matsuo}M. Matsuo, J. Ieda, K. Harii, E. Saitoh, and S. Maekawa, Phys. Rev. B {\bf 87}, 180402(R) (2013). 
\bibitem{SAW-SC:Hamada}M. Hamada, T. Yokoyama, and S. Murakami, Phys. Rev. B {\bf 92}, 060409(R) (2015).

\bibitem{Barnett1915}S. J. Barnett, Phys. Rev. {\bf 6}, 239 (1915).

\bibitem{Ono2015} M. Ono, H. Chudo, K. Harii, S. Okayasu, M. Matsuo, J. Ieda, R. Takahashi, S. Maekawa, and E. Saitoh, Phys. Rev. B {\bf 92}, 174424 (2015). 
\bibitem{Ogata2017} Y. Ogata, H. Chudo, M. Ono, K. Harii, M. Matsuo, S. Maekawa, and E. Saitoh, Appl. Phys. Lett. {\bf 110}, 072409 (2017).

\bibitem{Chudo2014} H. Chudo, M. Ono, K. Harii, M. Matsuo, J. Ieda, R. Haruki, S. Okayasu, S. Maekawa, H. Yasuoka, and E. Saitoh, Applied Physics Express {\bf 7}, 063004 (2014).  
\bibitem{Chudo2015} H. Chudo, K. Harii, M. Matsuo, J. Ieda, M. Ono, S. Maekawa, and E. Saitoh, J. Phys. Soc. Jpn. {\bf 84}, 043601 (2015).

\bibitem{Takahashi2016}R. Takahashi, M. Matsuo, M. Ono, K. Harii, H. Chudo, S. Okayasu, J. Ieda, S. Maekawa, and E. Saitoh, Nat. Phys. {\bf 12}, 52 (2016). 
See also, 
I. Zuti\'c and A. Matos-Abiague, Nat. Phys. {\bf 12}, 24 (2016);
D. Ciudad, Nat. Mater. {\bf 14}, 1188  (2015); 
J. Stajic, Science {\bf 20}, 924 (2015).

\bibitem{Bib:SpinConnection}D. Brill and J. Wheeler, Rev. Mod. Phys. {\bf 29} 465, (1957);  N. D. Birrell and P. C. W. Davies, {\it Quantum Fields in Curved Space}, (Cambridge University Press, Cambridge 1982).

\bibitem{FWT}L. L. Foldy and S. A. Wouthuysen, Phys. Rev. {\bf 78}, 29 (1950); S. Tani, Prog. Theor. Phys. {\bf 6}, 267 (1951).
\bibitem{LandauFluid}L. D. Landau and E. M. Lifshitz, {\it Fluid Mechanics}  (Pergamon, Oxford, 1987).


\end{thebibliography}
\end{document}